\documentclass[conference]{IEEEtran}
\usepackage{xcolor}
\usepackage{graphicx}
\usepackage[font=small,skip=0pt]{caption}
\usepackage{amsmath}
\usepackage{hyperref}
\usepackage{algorithm}
\usepackage[noend]{algpseudocode}
\hypersetup{colorlinks=true, pdfstartview=FitV, linkcolor=black, citecolor=black, urlcolor=blue}

\begin{document}
\title{Adaptive Lightweight Security for Performance Efficiency in Critical Healthcare Monitoring}
\author{\IEEEauthorblockN{Ijaz Ahmad\IEEEauthorrefmark{1}, Faheem Shahid\IEEEauthorrefmark{1}, Ijaz Ahmad\IEEEauthorrefmark{2}, Johirul Islam\IEEEauthorrefmark{1}, Kazi Nymul Haque\IEEEauthorrefmark{1}, Erkki Harjula\IEEEauthorrefmark{1} 
\IEEEauthorblockA{\IEEEauthorrefmark{1} Centre for Wireless Communications, University of Oulu, Finland, \\ \IEEEauthorrefmark{2}VTT Technical Research Centre of Finland, Finland} } }
\maketitle

\begin{abstract}
The healthcare infrastructure requires robust security procedures, technologies, and policies due to its critical nature. Since the Internet of Things (IoT) with its diverse technologies has become an integral component of future healthcare systems, its security requires a thorough analysis due to its inherent security limitations that arise from resource constraints. Existing communication technologies used for IoT connectivity, such as 5G, provide communications security with the underlying communication infrastructure to a certain level. However, the evolving healthcare paradigm requires adaptive security procedures and technologies that can adapt to the varying resource constraints of IoT devices. This need for adaptive security is particularly pronounced when considering components outside the ‘security sandbox of 5G’, such as IoT nodes and M2M connections, which introduce additional security challenges. This article brings forth the unique healthcare monitoring requirements and studies the existing encryption-based security approaches to provide the necessary security. Furthermore, this research introduces a novel approach to optimizing security and performance in IoT in healthcare, particularly in critical use cases such as remote patient monitoring. Finally, the results from the practical implementation demonstrate a marked improvement in the system performance.
\end{abstract}
\IEEEpeerreviewmaketitle

\section{Introduction}
In a rapidly evolving landscape of connectivity between IoT devices, machine-to-machine (M2M) protocols have played a pivotal role in facilitating efficient communication among devices. As we transition into an era of massive machine-type communications (mMTC), these M2M protocols continue to be instrumental, providing the foundation upon which mMTC is built. Among the several M2M protocols, the most common are the constrained application protocol (CoAP), which operates on the client-server model, and the message queueing telemetry transport (MQTT), which operates on the publish-subscribe model \cite{m2mprot}, \cite{secarchmqtt}. MQTT is a lightweight, publish-subscribe, message queuing protocol. It is an open Organization for the Advancement of Structured Information Standards (OASIS) standard and ISO recommendation (ISO/IEC 20922) and utilizes a broker, also known as a server, that hosts several topics. 

\begin{figure}
    \centering
    \includegraphics[width=0.8\linewidth]{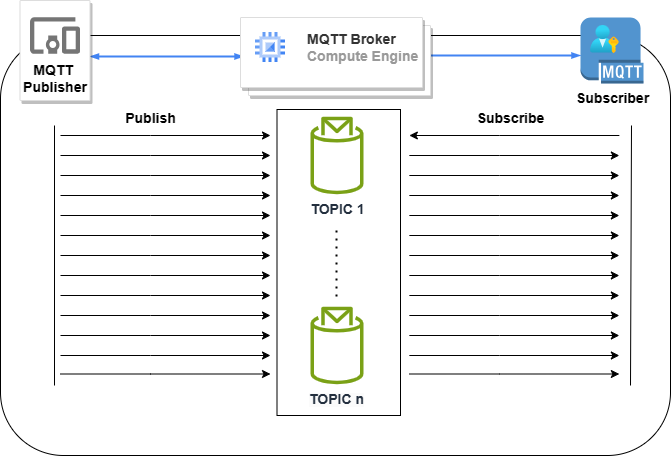}
    \setlength{\belowcaptionskip}{-5pt}
    \caption{MQTT architecture and protocol.}
    \label{mqttpubsub}
\end{figure}

A client can act as a publisher, sending data to the broker on a particular topic, and/or as a subscriber, receiving automated notifications whenever there is a new update on a topic they are subscribed to as depicted in Fig. \ref{mqttpubsub}. Although it is an OASIS standard, MQTT lacks basic security-related functionalities, such as mutual authentication, data confidentiality, and the integrity of transmitted data. The MQTT standard only supports one-way authentication, which relies on a pre-shared password for security purposes \cite{secarchmqtt}. However, due to the lack of a secure connection between the client (publisher/subscriber) and the MQTT broker, many security concerns related to authentication and confidentiality \cite{mqttsecreview}, \cite{secanalysisiot} have been highlighted lately. Existing solutions for MQTT such as Eclipse Mosquitto \cite{mosquitto2022eclipse}, EMQ \cite{EMQ}, and Apache ActiveMQ \cite{ApacheActiveMQ} include transport layer security (TLS) \cite{MQTTEncrypt} to provide a secure connection between each device and the MQTT broker.

This refers to the same security mechanisms that protect the privacy of HTTP transactions throughout the Internet. However, TLS is resource intensive, requires a substantial software framework, and is not ideal for devices with limited computational resources \cite{secarchmqtt}. MQTT, with its guaranteed delivery feature, is crucial for digital healthcare applications such as patient monitoring systems. This study emphasizes the need for end-to-end security. We explore the use of the lightweight cryptographic cipher ASCON, a NIST standard cipher, considering the latency requirements and resource constraints of the Internet of Medical Things (IoMT) in healthcare.

The main contributions of this article are as follows: First, we implement an experimental setup for remote patient monitoring use cases in healthcare care with MQTT as the underlying M2M protocol. Second, we analyze the performance overhead caused by TLS-based security in MQTT, considering the stringent security requirements of digital healthcare. Third, we investigate and explore the implementation of the NIST standard Advanced Encryption Standard (AES-128-GCM) and the ASCON lightweight cryptographic standard cipher suite for remote healthcare. Finally, we present performance enhancements with the use of the ASCON lightweight cryptographic cipher and its feasibility in future digital healthcare scenarios to ensure timely, secure and efficient communication from the incident site to the hospital.

The structure of the paper is as follows: Section II provides a comprehensive literature review, highlighting previous studies on MQTT in healthcare, security concerns, and the use of cryptographic algorithms. Section III details our research methodology, including our experimental setup, and our approach to measuring performance efficiency. In Section IV, we evaluate and discuss our findings, comparing the efficiency of various encryption schemes. We conclude the article in Section V, summarizing our research, its implications, and potential avenues for future research.

\section{Background and related work}
\subsection{MQTT for IoTs}
The MQTT protocol is specifically built to optimize bandwidth and minimize battery use in resource-constrained Internet of Things (IoT) devices while providing guaranteed delivery of messages. Therefore, it is preferred to UDP-based CoAP \cite{m2mprot} since it is based on the Transmission Control Protocol (TCP) and ensures communication reliability with flexibility by offering the following three levels of Quality of Service (QoS).\\
\textbf{\textit{QoS0 - Fire and forget (at most once)}}: A message is transmitted just once without the need for any acknowledgment.\\
\textbf{\textit{QoS1 - At least once}}: A message is transmitted at least once, and confirmation is necessary. In case of no confirmation, the message is retransmitted repeatedly.\\
\textbf{\textit{QoS2 - Exactly once}}: It employs a four-way handshake approach to ensure that the message arrives at the recipient just once.

MQTT QoS1 can be used in latency-sensitive applications such as digital healthcare where QoS0 fails to provide acknowledgment and QoS2 is too slow to meet the latency requirements of digital healthcare use cases. MQTT, although using TCP, is specifically designed to minimize overhead compared to other TCP-based application layer protocols \cite{perfcomp}.

\subsection{MQTT in healthcare}
\begin{figure}[h] 
   \centering
   \includegraphics[width=\linewidth]{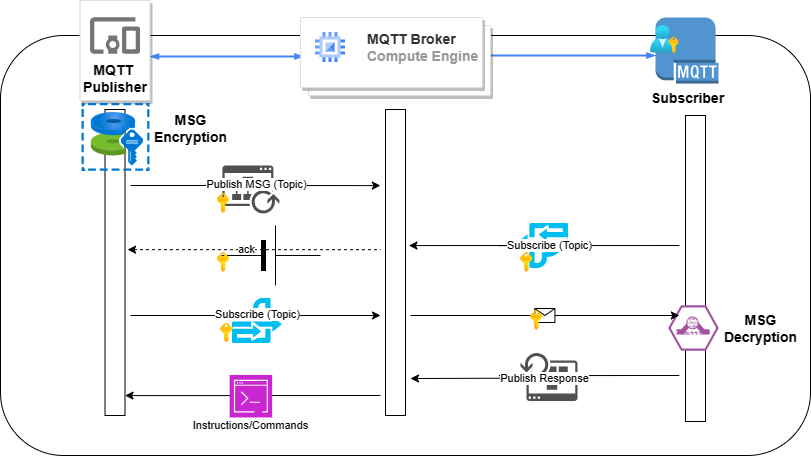}
   \setlength{\belowcaptionskip}{-5pt}
   \caption{MQTT (QoS 1) Communication Sequence Diagram.}
 \label{sequencediag}
 \end{figure}

MQTT can be deployed in remote patient monitoring healthcare scenarios to communicate patient's vital information such as heart rate, body temperature, and patient movement information \cite{smarthealthmqtt}. This helps doctors monitor their patients from anywhere and anytime and provides an opportunity for patients' relatives to observe their health conditions remotely. Minimizing the communication latency and securing the data exchanged are the prominent features of healthcare systems. \cite{noderedmqtt} discusses a novel methodology that may be used to accurately measure latency in data exchange between MQTT publishers and subscribers and recommends QoS 1 as the most favorable balance between achieving minimal end-to-end latency and minimizing unpredictability. To secure MQTT data exchange, MQTT QoS 1 can be used with symmetric encryption between MQTT clients, as shown in Fig. \ref{sequencediag}.

\subsection{Security in MQTT}
The default configuration of the MQTT protocol uses no encryption for messages exchanged among the broker and clients \cite{m2msurveymqtt}, which makes it not acceptable for health data, since the confidentiality of the data exchanged is a core requirement in IoMT. The security and performance of the MQTT protocol are correlated inversely \cite{mqtttlsperf} and the choice of having security impacts the performance of the protocol. This study offers significant information on the consequences of including security measures in MQTT and emphasizes the need to adopt a well-balanced strategy to ensure both security and performance. 

Compared to the conventional TLS security setting, \cite{MQTTEncrypt} suggests the use of Fernet cipher for encryption in MQTT. The MQTT security survey \cite{nwiah2021survey} emphasizes the need for lightweight security and suggests Hummingbird-2, as an ultra-lightweight cryptography, based on the issues identified in the survey. \cite{mqttencryptiot} discovered XXTEA cipher to be efficient after implementing AES128, ECC, ChaCha20 and XXTEA in the MQTT protocol with the ESP8266 IoT chip. Recently, \cite{avery2022analysis} presented an innovative assessment of the ASCON lightweight cryptographic method in practical scenarios. The study reveals that ASCON can effectively encrypt messages in communication, ensuring both confidentiality and integrity of the data exchanged. 

In the context of Remote Patient Monitoring (RPM) in healthcare, it is crucial to ensure both data confidentiality and efficient real-time communication with minimal communication latency. Adding a security mechanism to the standard MQTT protocol introduces overhead and hence limits its performance efficiency \cite{mqtttlsperf, nwiah2021survey, mqttencryptiot}. With ASCON’s recent recognition as a NIST standard for lightweight cryptography \cite{avery2022analysis}, it becomes a potential candidate for RPM in specific and healthcare in general in combination with the lightweight M2M protocol, MQTT. This research aims to explore this potential by investigating the trade-off between security and efficiency and implementing ASCON-128 in MQTT for the RPM use case in healthcare.

\section{Research Methodology}
\subsection{Use-Case: Remote Patient Monitoring in Healthcare}
\begin{figure}[ht]
  \centering
  \includegraphics[width=\linewidth]{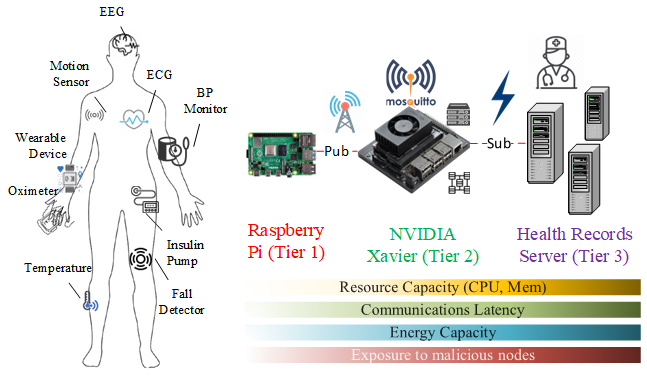}
  \caption{Remote Patient Monitoring (RPM) - Use case Diagram.}
  \label{usecasediag}
\end{figure}

We focus on the Remote Patient Monitoring (RPM) use case in Healthcare due to its critical nature that requires adaptive security procedures \cite{ahmad2024adaptive}. The use-case scenario involves the deployment of multiple sensors to continuously monitor an emergency patient. This monitoring begins at the incident site and continues during the patient’s transfer to the hospital. The use case diagram is presented in Fig. \ref{usecasediag}. Since resource-constrained devices publishing the data may not connect directly to the broker when SSL-based TLS is used, an intermediary node may be placed between them \cite{wardana2018access}. In our study, we used a single-board computer (Raspberry Pi) in this role. In a real-life scenario, this node could be a mobile phone, a tablet computer, or a dedicated paramedic terminal. This scenario is particularly relevant in critical care situations where continuous patient monitoring is essential for timely and effective medical intervention. We divide the remote patient monitoring use case into the following steps:

\textbf{Data Collection:} Multiple IoMT sensors are attached to the patient at the incident site. These sensors are set to continuously monitor various vital signs such as heart rate, blood pressure, oxygen saturation, etc. The data from these sensors is collected by a Raspberry Pi device that serves as a local edge computing node.

\textbf{Data Transmission:} The RPi device is equipped with a 5G modem, enabling high-speed and low-latency communication. It is capable of publishing the collected data from the IoMT sensors to a broker using the MQTT protocol, which is known for its light overhead and real-time capabilities.

\textbf{Data Processing and Analysis:} The MQTT broker is configured on a GPU-powered NVIDIA Jetson Xavier NX device, serving as an edge server. Jetson Xavier’s edge AI capabilities, scalability \cite{xaviercluster}, and real-time processing make it suitable as an edge server for our use case. It also offers a cost-effective solution for experimental setups or small-scale deployments. Future work could explore comparisons with other MEC solutions that align more closely with ETSI standards \cite{etsi2019multi}. This device receives the published data, processes it, and can trigger alerts or actions based on the data received.

\textbf{Security:} To ensure the privacy and security of sensitive health data, the communication between clients and the broker is encrypted. We evaluate the performance of candidate encryption schemes. The MQTT broker receives the published information and makes it available to subscribing entities on a particular topic. These could include emergency medical services, doctors in the hospital, or a medical data processing system. The MQTT clients have a pre-shared secret key (or X.509 certificates) to maintain confidentiality in the system. Through this use case, our objective is to demonstrate the feasibility and efficiency of using MQTT in healthcare, particularly in scenarios that require remote patient monitoring in real time. The performance and security trade-offs of using different encryption methods in this context are also analyzed and discussed.
\subsection{The Experimental Setup}
Evaluating the impact of security methods in real-world situations is crucial as it helps in identifying possible complications and the overhead caused. Furthermore, it may demonstrate the protocol's ability to handle scalability in effectively managing a substantial volume of messages. An experimental setup has been created to evaluate the efficacy of the NIST standard lightweight cryptographic cipher ASCON, compared to the NIST standard block cipher AES and MQTT native TLS, encrypting health data for the remote patient monitoring healthcare use case using MQTT as an M2M communication protocol. Our setup involves a real-time experiment, using the Python IDE, supported libraries, Paho MQTT client, and the Mosquitto broker. Detailed software specifications along with the usage purpose are listed in Table \ref{tab:swspecs}.

\begin{table}[ht]
\caption{Software Specifications}
\centering
\begin{tabular}{|p{1cm}|p{4cm}|p{2.5cm}|}
\hline
\textbf{Software / Library} & \textbf{Purpose} & \textbf{Configuration} \\
\hline
Python & v3.8, Establishing a connection, controlling communication, implementing encryption and decryption algorithms & pre-installed on all devices \\
\hline
Paho-mqtt client & Allows devices to connect to an MQTT broker. Encrypt and publish the messages, receive and decrypt the messages & configure on MQTT publisher and subscriber nodes. \\
\hline
ASCON & Lightweight cryptographic cipher for payload encryption using ASCON-128. & pyAscon library is used for ASCON cipher. \\
\hline
Crypto & python library that provides cryptographic recipes and primitives and is used for the implementation of AES. & AES encryptor and decryptor configured with key and message.\\
\hline
NumPy & Supports large, multi-dimensional arrays and is used for storing mean round-trip times for encryption and decryption & Configured with .csv file exports for results. \\
\hline
Matplotlib & Provides a plotting framework and is used along with NumPy to analyze the round-trip times recorded visually. & Visualization is done as a combo graph with .csv files. \\
\hline
Mosquitto & Open Source message broker that implements the MQTT protocol. & Configured on Xavier with mosquitto.conf file. \\
\hline
Wireshark & Powerful tool for capturing and analyzing network traffic in real time. & Configured with promiscuous mode on Eavesdropper host. \\
\hline
\end{tabular}
\label{tab:swspecs}
\end{table}

\begin{table}[ht]
\caption{Hardware Specifications}
\centering
\begin{tabular}{|p{1cm}|p{5cm}|p{1.5cm}|}
\hline
\textbf{Device} & \textbf{Specifications} & \textbf{Role} \\
\hline
Raspberry Pi 3 Model B & CPU: Broadcom BCM2837B0 quad-core A53 (ARMv7) 32-bit @ 1.4GHz, Raspberry Pi OS RAM: 1GB LPDDR2, Storage: microSD Networking: 10/100 Ethernet, 2.4GHz 802.11n Bluetooth: Bluetooth 4.1 Classic, BLE & Publisher and Subscriber \\
\hline
NVIDIA Jetson Xavier NX & CPU: 6-core Carmel ARM®v8.2 64-bit, cache: 6MB (L2) + 4MB (L3), Storage: microSD GPU: NVIDIA Volta architecture with 384 CUDA cores and 48 Tensor cores, OS: Ubuntu 18.04 RAM: 8 GB 128-bit LPDDR4x @ 51.2GB/s & Broker\\
\hline
Quectel 5G Modem & Model: Quectel RM500Q-GL Frequency Bands: Supports 5G NR Sub-6 GHz, LTE-A and WCDMA Data Rate: 5G NR up to 4.0 Gbps (DL), 500 Mbps (UL) Interfaces: USB 3.1, PCIe 3.0, UART, GPIO etc. Operating Temperature: -40°C to +85°C & Comm Interface \\
\hline
Windows Laptop & Intel-based laptop with Windows 11 OS configured with eavesdropping tools for monitoring the traffic in the network & Eavesdropper \\
\hline
\end{tabular}
\label{tab:hwspecs}
\end{table}

\begin{figure} 
  \centering
  \includegraphics[width=0.9\linewidth]{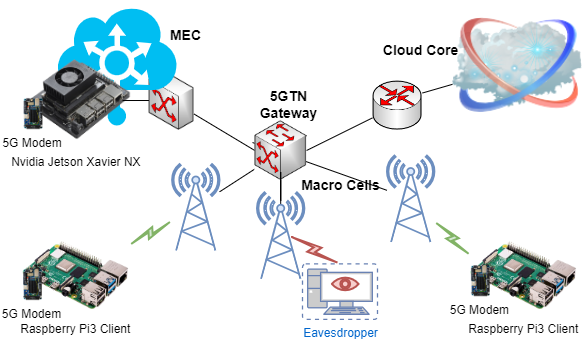}
  \setlength{\belowcaptionskip}{-5pt}
  \caption{Network Topology Diagram.}
\label{commdiag}
\end{figure}

The experiment is designed to mimic a real-world scenario of multi-sensor monitoring of an emergency patient from the incident site to the hospital. The Raspberry Pi, acting as a local edge device, collects patient information and publishes it to an MQTT broker configured on an NVIDIA Jetson Xavier, which serves as an edge server. Detailed hardware specifications are listed in Table \ref{tab:hwspecs}. Client communication is facilitated by a Quectel 5G modem, connecting to the University of Oulu’s 5G Test Network. The 5G Test Network (5GTN) is an accessible experimental platform for the next-generation mobile communication networks, offering opportunities for service creators and network operators to explore innovative solutions \cite{5GTN_2019}. The 5G Test Network (5GTN) at the University of Oulu consists of a Radio Access Network (RAN) that operates on permitted LTE and 5G frequency bands \cite{5GTN_2019}. The network topology diagram is illustrated in Fig. \ref{commdiag}.

\subsection{Test Cases}
We have prepared a series of test cases to compare the performance of no cipher (no encryption), AES, TLS, and ASCON in this context. We have considered the following test cases in this experiment to provide valuable information on the performance and security trade-offs of different encryption methods in an MQTT-based remote patient monitoring system.
\subsubsection{No Cipher} In this case, MQTT messages are sent as \textit{plaintext} without encryption scheme. This serves as a baseline for comparison with the other test cases. It allows us to understand the overhead introduced by the encryption process in terms of time and computational resources.
\subsubsection{AES-GCM-128} This test case involves the implementation of AES with a 128-bit key to encrypt the MQTT messages. AES-128 is widely used due to its balance of security and performance. GCM is a mode of AES that uses the Counter (CTR) mode to encrypt data and uses Galois mode for authentication \cite{aesgcm}. This test case will help us understand how a conventional encryption method performs in our setup.
\subsubsection{ASCON-128} In this test case, we use ASCON, a lightweight cryptographic cipher designed for efficiency in constrained environments. With a 128-bit key, ASCON-128 provides a high level of security while being resource-friendly. This test case will provide insights into the performance trade-offs of using lightweight ciphers in MQTT.
\subsubsection{SSL/TLSv1.2} This test case involves the use of TLS version 1.2 to secure MQTT messages. TLSv1.2 provides robust security and is widely adopted in various MQTT applications such as Eclipse Mosquitto \cite{mosquitto2022eclipse}, EMQ \cite{EMQ}, and Apache's ActiveMQ \cite{ApacheActiveMQ}. By comparing this with the other test cases, we can evaluate the suitability of using such a comprehensive security protocol in a remote patient monitoring scenario.

\section{Results and Evaluation}
In this section, we delve into the outcomes of our real-time experiment conducted in the Python development environment. Leveraging the power of associated libraries, we implemented the test cases defined in the previous section. In all our test scenarios, we utilized test vectors as presented in Table \ref{vectorEncrypt}. A \textit{plaintext} message composed of repeated instances of the letter ‘X’, the repetition frequency being contingent on the message size. We employed a secret key and an Initialization Vector (IV) / Nonce, both of 16 bytes (128 bits), for the AES and ASCON test cases. For ASCON, we also incorporated an \textit{associated data} text of 16 bytes (128 bits). 

In our experiment, we meticulously measured the round-trip time for the journey of a message through our MQTT setup. This journey begins at the publisher node, where the message is first encrypted. From there, it is transmitted to the broker, and subsequently received by the subscriber node. At the subscriber node, the message is decrypted, marking the completion of its round-trip. To be able to measure the time elapsed precisely, the Raspberry Pi client is made as both the subscriber and publisher. The time library in Python is used to record the CPU time before encryption till the message is decrypted after successful receipt at the subscriber node. Our evaluation Algo. \ref{evalalgo} applies to all test cases, ensuring consistency and fairness.

\begin{algorithm}
\caption{Mean Round Trip Time calculation for MQTT}
\label{evalalgo}
\begin{algorithmic}[1]
\State Establish a connection to the Broker with a specific topic.
\State Self-subscribe to the topic.
\State Collect the input from the IoMT sensors.
\State Initialize $MRTT$
\For{$2^N$ iterations, $N$=(1,2,..., 12)}
    \For{$count =$ 1 to 10}
        \State Start the timer.
        \State Encrypt the message.
        \State Publish the encrypted message to the broker.
        \State Wait to receive the update from the Broker.
        \State Decrypt the received message.
        \State Stop the timer.
        \State Calculate round-trip time $RTT$.
    \EndFor
    \State Calculate $MRTT$.
\EndFor
\State Disconnect from the broker and return $MRTT$.
\end{algorithmic}
\end{algorithm}

\begin{figure}
    \centering
    \includegraphics[width=\linewidth]{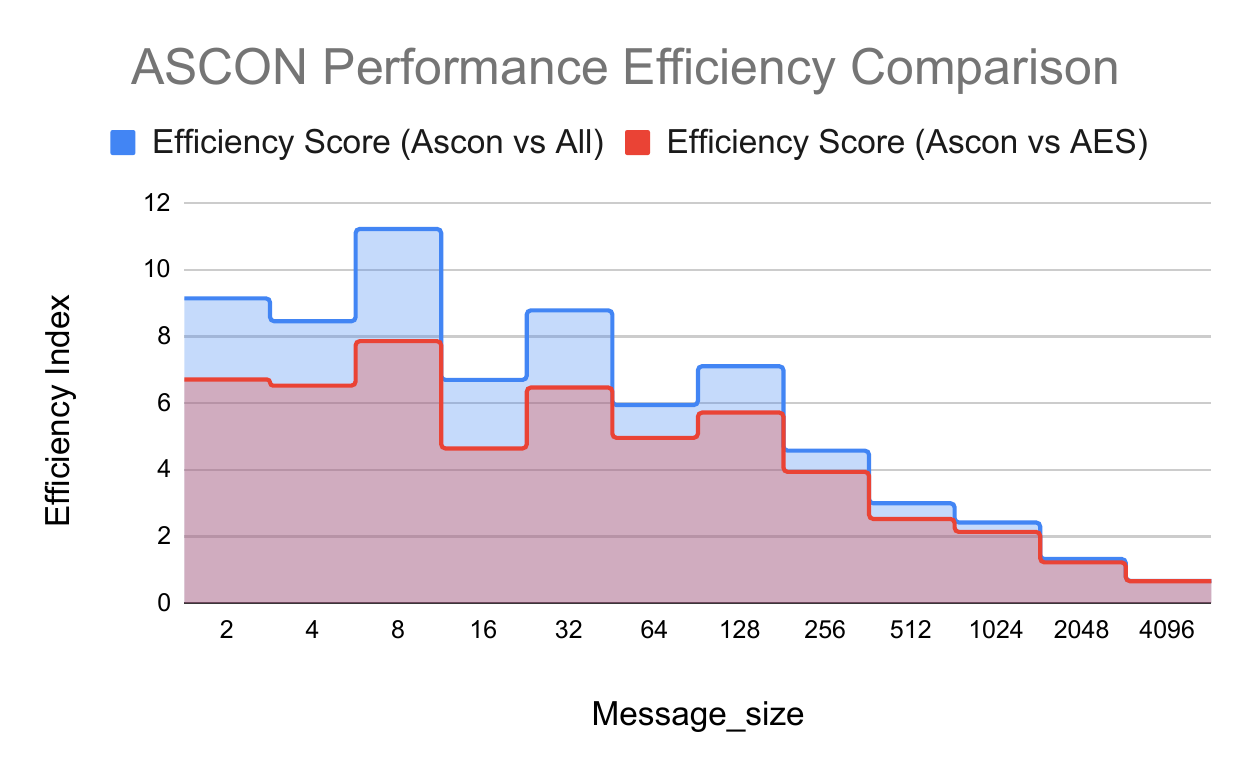}
    \vspace*{-5mm}
    \setlength{\belowcaptionskip}{-10pt}
    \caption{Performance Efficiency of ASCON.}
    \label{fig:asconPerf}
\end{figure}

Following the algorithm, each round doubles $message\_size$ and encompasses ten (10) iterations that comprise encryption, transmission, and decryption processes. Fig. \ref{fig:asconPerf} reflects the performance efficiency of ASCON-lightweight cipher when compared with all others, and AES being the closest to ASCON in performance efficiency. The vertical axis represents the Performance Efficiency Index, which is defined as the comparison of the ASCON MRTT value to the mean MRTT of other ciphers. The key findings from the evaluation results are discussed below.

\begin{figure}[ht]
  \centering
  \includegraphics[width=\linewidth]{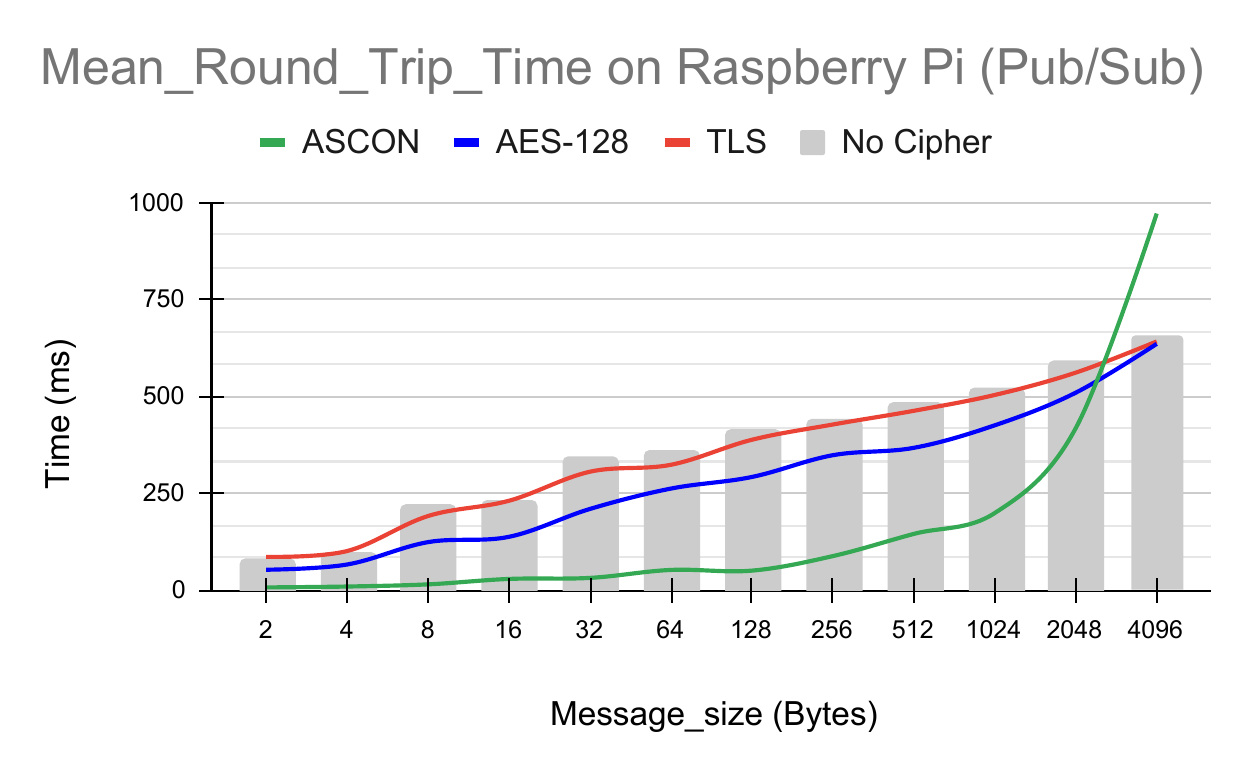}
  \vspace*{-5mm}
  \setlength{\belowcaptionskip}{-10pt}
  \caption{Cipher Performance Evaluation on Raspberry Pi.}
\label{Pi}
\end{figure}

\begin{figure}[ht]
  \centering
  \includegraphics[width=\linewidth]{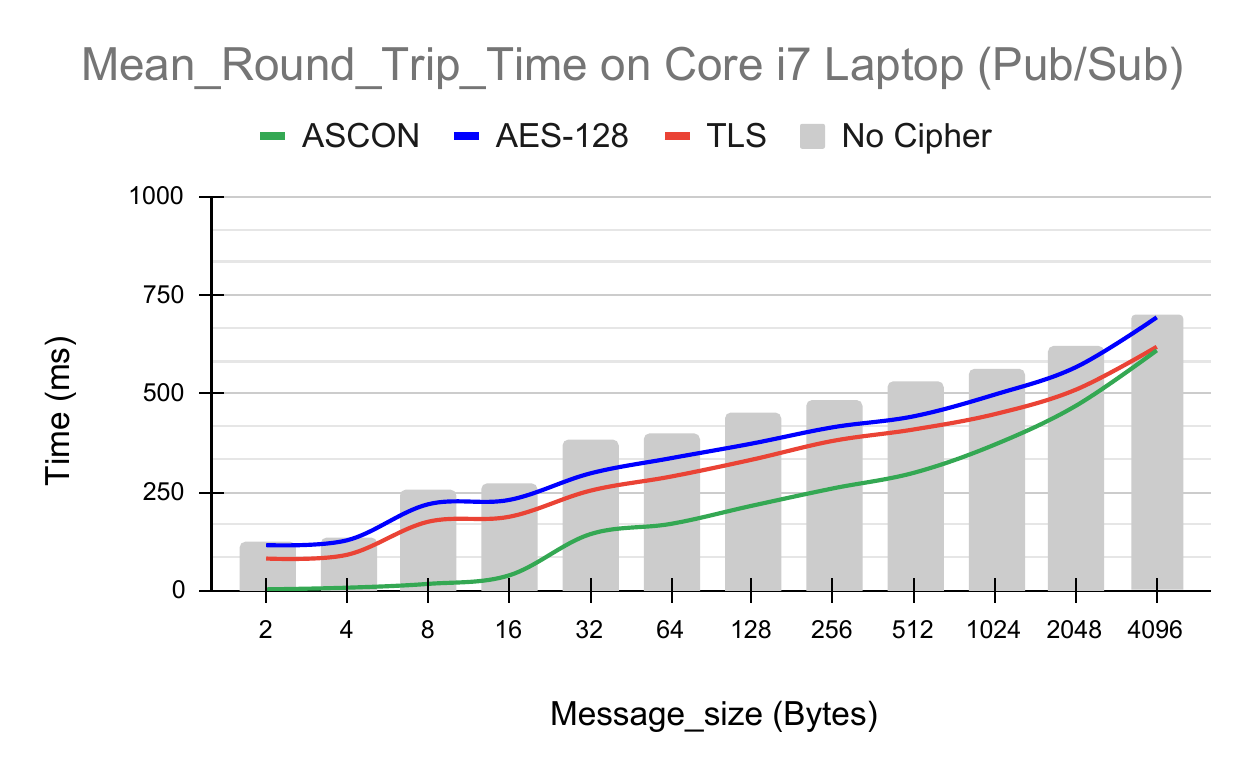}
  \vspace*{-5mm}
  \setlength{\belowcaptionskip}{-10pt}
  \caption{Cipher Performance Evaluation on Laptop.}
  \label{laptop}
\end{figure}

The results presented in this section provide a comprehensive analysis of the performance trade-offs between the lightweight cryptographic cipher ASCON, AES, and TLS in MQTT for the remote patient monitoring healthcare use case.  The round-trip time considers various underlying factors such as computational overhead of encryption and decryption scheme, communication latency of 5G network, MQTT publish/subscribe model efficiency, and data throughput, providing a holistic comprehension of the security and performance trade-offs involved. Fig. \ref{laptop} represents values when a laptop is used as publisher and subscriber, while Fig. \ref{Pi} represents values when the implementation is done on a Raspberry Pi 3 node.

For each test case, the mean round-trip times are visualized in Figs. \ref{fig:asconPerf}, \ref{Pi}, and \ref{laptop}. Figs. \ref{Pi} and \ref{laptop} depict the round trip time in milliseconds for MQTT (QoS 1) with different encryption schemes as the message size in bytes increases. The x-axis represents the message size in bytes, ranging from 0 to 4096 ($2^{12}$ = 4KB), while the y-axis represents the time in milliseconds, ranging from 0 to 1000 (one second). In both figures. \ref{Pi} and \ref{laptop} for all ciphers, the round trip time increases, as the message size increases. This is expected as larger messages take more time to transmit and process.

\subsection{} 
\subsubsection{\textbf{Symmetric results on laptop and Raspberry Pi}} As depicted in Figs. \ref{Pi} and \ref{laptop}, both graphs are closely symmetric, and there is no significant behavioral deviation since the software development environment and algorithm source codes are the same in both cases. The only noticeable difference is the for message\_size 4096, where the MRTT for Ascon is exponentially raised and confirmed with multiple iterations. A similar pattern is observed in the laptop Fig. \ref{laptop}, indicating a sharp rise with the increase in the message size.

\subsubsection{\textbf{Baseline case (No cipher)}} The columns in Figs. \ref{Pi} \& \ref{laptop} represent the baseline round trip time for MQTT communication. This represents the mean round-trip time without encryption. While this test case might not seem significant on its own, it serves as a baseline for comparing the performance of other encryption schemes. Interestingly, in Fig. \ref{laptop}, this case does not show the shortest round-trip time for message sizes smaller than 4 KB, which is surprising from what one might expect. This could be attributed to various factors such as network conditions, Quality of Service (QoS) level, server load, or client performance. Another possible explanation is that the encryption schemes utilized require data to be processed in 128-byte blocks, which could lead to more efficient processing for smaller data sizes. As the message size increases, the round trip time tends to normalize and improves for message sizes greater than 4 KB.

\subsubsection
{\textbf{ASCON is performance-efficient for smaller messages}} As depicted in Fig. \ref{fig:asconPerf}, ASCON-128 exhibits superior performance when compared to others. As illustrated in Fig. \ref{fig:asconPerf} For performance comparison, for a smaller message size of 8 bytes, ASCON is approximately 11x (times) better than others and 8x (times) efficient than AES. For a message size of 1024 bytes (1 KB), ASCON is twice as efficient as the others. Performance is close to that of others for a message size of 2048 bytes (~2 KB) but still marginally better. This suggests that ASCON-128 strikes an optimal balance between security and performance for smaller messages. Similarly, as shown in Fig. \ref{fig:asconPerf}, ASCON-128’s performance surpasses others across the entire graph when evaluated on Raspberry Pi or Laptop.

\subsubsection{\textbf{Cipher efficiency and larger messages}} It is not a rule of thumb that lightweight ciphers will always surpass conventional and transient security mechanisms. For example, in this experiment, we have observed that when the message size exceeds 2048 Byes (~2 KB), the lightweight encryption standard ASCON starts to degrade as shown in Fig. \ref{Pi} on Raspberry Pi with MQTT. In the remote patient monitoring use case, messages include patient vitals that are not bulky, but message size needs to be considered when deciding on the selection of an encryption algorithm. 

\subsubsection{\textbf{Performance of AES-128 and TLSv1.2}} In both Figs. \ref{Pi} and \ref{laptop}, the AES-128 and TLSv1.2 lines closely align, indicating similar mean round trip times. This is expected as native TLSv1.2 intrinsically uses the same AES cipher with a 128-bit key size. However, TLSv1.2 can be operationally complex due to the need for generating and distributing X.509 certificates, especially as the number of clients increases. Notably, TLSv1.2 performs marginally better on a laptop than on Raspberry Pi, while AES-128-GCM shows superior performance on the Raspberry Pi client.


\begin{table}[ht]
\caption{Encryption Test Vectors}
\centering
\begin{tabular}{|l|l|}
\hline
\textbf{Secret Key} & 'This is 16 bytes' \\ \hline
\textbf{IV/Nonce} & 'This is 16 bytes' \\ \hline
\textbf{Associated Data} & 'This is 16 bytes' \\ \hline
\textbf{Plaintext} & 'XX' (for message\_size 2, repeated for others) \\ \hline
\textbf{Ciphertext} & depending on cipher \\ \hline
\textbf{Message\_size} & 2, 4, 8, 16, 32, 64, 128, 256, 512, 1024, 2048, 4096 \\ \hline
\end{tabular}
\label{vectorEncrypt}
\end{table}

Given the typical message sizes in remote patient monitoring use cases, generally less than 2 KB, ASCON-128 emerges as a superior and more applicable choice. Its exceptional performance for smaller messages, as demonstrated in our evaluations, indicates an optimal balance between security and performance. This makes ASCON-128 particularly well suited for remote patient monitoring scenarios where efficient data transmission is crucial without compromising security.

\section{Conclusion and Future Work}
This study has highlighted the critical need for flexible and efficient security measures within the dynamic landscape of healthcare. We emphasize the importance of lightweight and adaptive security schemes, particularly in the context of IoT devices and M2M connections that operate beyond the protective confines of the 5G security framework. Our study establishes ASCON-128 as a performance-efficient cipher for smaller messages, significantly outperforming others. However, it is important to note that lightweight ciphers like ASCON may not always surpass conventional security mechanisms, especially for larger messages. 

Our real-time experiment, conducted on the 5G Test Network (5GTN), provided practical insight into the performance trade-offs involved. Compared to other techniques, our approach offers a more adaptable solution, capable of adapting to the varying resource constraints of IoT devices. This adaptability makes it particularly suitable for healthcare IoT applications, where device resources can be highly variable. The research findings contribute to ongoing efforts to optimize security and performance in healthcare IoT, thus paving the way for a more secure and reliable healthcare infrastructure.

Looking ahead, the research carried out in this work will be extended to a Proof of Concept (PoC) environment and tested through the commercial 5G network. As AI and Machine Learning (ML) become more seamlessly integrated with 6G technology, we anticipate significant advances in real-time adaptivity. These innovations will enable decision-making processes with minimal human intervention, particularly in critical healthcare monitoring scenarios. This represents an exciting direction for future research and development in adaptive security for healthcare with IoTs.

\section*{Acknowledgment}
This research is supported by the Business Finland projects Eware-6G (grant 8819/31/2022), Tomohead (grant 8095/31/2022), SUNSET-6G (grant 8682/31/2022), and the Research Council of Finland 6G Flagship program (grant number 346208).

\bibliographystyle{unsrt}
\bibliography{IEEEtran5/ref}

\begin{thebibliography}{10}

\bibitem{m2mprot}
Biswajeeban Mishra and Attila Kertesz.
\newblock The use of mqtt in m2m and iot systems: A survey.
\newblock {\em IEEE Access}, 8:201071--201086, 2020.

\bibitem{secarchmqtt}
Chang-Seop Park and Hye-Min Nam.
\newblock Security architecture and protocols for secure mqtt-sn.
\newblock {\em IEEE Access}, 8:226422--226436, 2020.

\bibitem{mqttsecreview}
Fu~Chen, Yujia Huo, Jianming Zhu, and Dan Fan.
\newblock A review on the study on mqtt security challenge.
\newblock In {\em 2020 IEEE International Conference on Smart Cloud (SmartCloud)}, pages 128--133, 2020.

\bibitem{secanalysisiot}
Jos{\'e} Rold{\'a}n-G{\'o}mez, Javier Carrillo-Mond{\'e}jar, Juan~Manuel Castelo~G{\'o}mez, and Sergio Ruiz-Villafranca.
\newblock Security analysis of the mqtt-sn protocol for the internet of things.
\newblock {\em Applied Sciences}, 12(21):10991, 2022.

\bibitem{mosquitto2022eclipse}
Eclipse Mosquitto.
\newblock Eclipse mosquitto™ an open source mqtt broker.
\newblock {\em URL: https://mosquitto. org}, 2022.

\bibitem{EMQ}
Emq—the massively scalable mqtt broker for iot and mobile applications.
\newblock \url{http://emqtt.io/}.
\newblock Accessed: 2024-01-12.

\bibitem{ApacheActiveMQ}
Activemq—flexible \& powerful open source multi-protocol messaging.
\newblock \url{http://activemq.apache.org/apollo/}.
\newblock Accessed: 2023-12-15.

\bibitem{MQTTEncrypt}
Suja~P Mathews and Raju~R Gondkar.
\newblock Protocol recommendation for message encryption in mqtt.
\newblock In {\em 2019 International Conference on Data Science and Communication (IconDSC)}, pages 1--5, 2019.

\bibitem{perfcomp}
Manasi Mishra and S.~R. Reddy.
\newblock Performance assessment and comparison of lightweight d2d-iot communication protocols over resource constraint environment.
\newblock {\em Multimedia Tools and Applications}, 2024.

\bibitem{smarthealthmqtt}
Borade~Samar Sarierao and Amara Prakasarao.
\newblock Smart healthcare monitoring system using mqtt protocol.
\newblock In {\em 2018 3rd International Conference for Convergence in Technology (I2CT)}, pages 1--5, 2018.

\bibitem{noderedmqtt}
Akshatha~P S and S~M~Dilip Kumar.
\newblock Delay estimation of healthcare applications based on mqtt protocol: A node-red implementation.
\newblock In {\em 2022 IEEE International Conference on Electronics, Computing and Communication Technologies (CONECCT)}, pages 1--6, 2022.

\bibitem{m2msurveymqtt}
Biswajeeban Mishra and Attila Kertesz.
\newblock The use of mqtt in m2m and iot systems: A survey.
\newblock {\em IEEE Access}, 8:201071--201086, 2020.

\bibitem{mqtttlsperf}
A.~R. Alkhafajee, Abbas M.~Ali Al-Muqarm, Ali~H. Alwan, and Zaid~Rajih Mohammed.
\newblock Security and performance analysis of mqtt protocol with tls in iot networks.
\newblock In {\em 2021 4th International Iraqi Conference on Engineering Technology and Their Applications (IICETA)}, pages 206--211, 2021.

\bibitem{nwiah2021survey}
Edward Nwiah and Shri Kant.
\newblock A survey on securing payload in mqtt and a proposed ultra-lightweight cryptography.
\newblock In {\em Information and Communication Technology for Intelligent Systems: Proceedings of ICTIS 2020, Volume 2}, pages 323--335. Springer, 2021.

\bibitem{mqttencryptiot}
Yasir Iqbal, Muhammad~Faisal Amjad, Fawad Khan, and Haider Abbas.
\newblock The implementation of encryption algorithms in mqtt protocol for iot constrained devices.
\newblock In {\em 2022 14th International Conference on Computational Intelligence and Communication Networks (CICN)}, pages 804--810, 2022.

\bibitem{avery2022analysis}
J~Avery, B~Fraelich, W~Duran, A~Lee, A~Sullivan, Z~Mechalke, MB~Birrer, S~Dick, and J~Cochran.
\newblock Analysis of practical application of lightweight cryptographic algorithm ascon.
\newblock {\em Computer Security Resource Center}, 2022.

\bibitem{ahmad2024adaptive}
Ijaz Ahmad, Ijaz Ahmad, and Erkki Harjula.
\newblock Adaptive security in 6g for sustainable healthcare.
\newblock {\em arXiv preprint arXiv:2403.01100}, 2024.

\bibitem{wardana2018access}
Aulia~Arif Wardana and Riza~Satria Perdana.
\newblock Access control on internet of things based on publish/subscribe using authentication server and secure protocol.
\newblock In {\em 2018 10th International Conference on Information Technology and Electrical Engineering (ICITEE)}, pages 118--123. IEEE, 2018.

\bibitem{xaviercluster}
Steve McIntire.
\newblock Building a four-node cluster with nvidia jetson xavier nx , [online] available: https://developer.nvidia.com/blog/building-a-four-node-cluster-with-nvidia-jetson-xavier-nx/, Jul 2022.

\bibitem{etsi2019multi}
M~Etsi.
\newblock Multi-access edge computing (mec) framework and reference architecture.
\newblock {\em ETSI GS MEC}, 3:V2, 2019.

\bibitem{5GTN_2019}
Explore 5g features and performance in a controlled, real environment, [online] available: https://5gtn.fi/, Mar 2019.

\bibitem{aesgcm}
Byung-Yoon Sung, Ki-Bbeum Kim, and Kyung-Wook Shin.
\newblock An aes-gcm authenticated encryption crypto-core for iot security.
\newblock In {\em 2018 International Conference on Electronics, Information, and Communication (ICEIC)}, pages 1--3, 2018.

\end{thebibliography}
\end{document}